# Direct measurement of the intermolecular forces confining a single molecule in an entangled polymer solution

Rae M. Robertson and Douglas E. Smith

*Department of Physics, University of California, San Diego*
*Mail Code 0379, 9500 Gilman Drive, La Jolla, CA 92093*
Email: des@physics.ucsd.edu, rroberts@physics.ucsd.edu

June 22, 2007

We use optical tweezers to directly measure the intermolecular forces acting on a single polymer imposed by surrounding entangled polymers (115 kbp DNA, 1 mg/ml). A tube-like confining field was measured in accord with the key assumption of reptation models. A time-dependent harmonic potential opposed transverse displacement, in accord with recent simulation findings. A tube radius of 0.8 μm was determined, close to the predicted value (0.5 μm). Three relaxation modes (~0.4, 5 and 30 s) were measured following transverse displacement, consistent with predicted relaxation mechanisms.

Over the past several decades much effort has been directed at understanding the physical properties of entangled polymer solutions and melts. The reptation theory introduced by P. G. de Gennes and extended by Doi and Edwards and others has proven quite successful in describing many experimental findings [1-4]. The key assumption of the theory is that on short time scales each molecule is confined, due to entanglements with surrounding molecules, to move within a tube-like region parallel to its own contour. Motion transverse to the molecular contour is predicted to be highly restricted compared with parallel motion. The advantage of the tube model is that it reduces a complex many-body problem to that of a single polymer moving in an effective mean field. While originally developed for melts, tube models have been successfully extended to polymer solutions using "blob" theory [1, 5], in which each chain is divided into correlation blobs of length $\zeta$. Thus, the solution becomes a melt of chains with effective monomer size $\zeta$ [6].

While the tube theory has proven quite useful, the notion of a "tube-like constraint" has remained rather qualitative. No experiment has directly measured the intermolecular forces acting on a single entangled polymer to oppose its displacements. An expression for the tube radius *a* has been predicted theoretically [2, 7], but in practice it has only been defined empirically via measurements of the bulk plateau modulus [2, 8]. Thus, the confining tube is present in most theories as an assumption rather than being derived from fundamental molecular properties. Elucidation of the nature of this entanglement field has been identified as one of the primary open challenges in the field [4]. Recently Zhou and Larson have presented a theoretical method for directly calculating the tube potential in a melt via molecular dynamics simulations [9].

In previous work, we directly observed tube-like motion of single entangled DNA molecules by fluorescence microscopy [10, 11], however we did not quantify the confining forces. A simple picture of the transverse confining potential would be a "hard-walled" square well of radius $a$ restricting displacement. However, individual entanglements certainly lie at distances both smaller and larger than $a$, and the surrounding polymers are dynamic and able to reptate, stretch, and deform, causing individual constraints to constantly disappear, reappear, and change locations [4, 12]. In their recent simulations, Zhou and Larson calculated that the tube radius actually increases gradually with time [9].

Behavior consistent with predictions of tube models has been observed in bulk rheology experiments and these data have been analyzed to infer the properties of the tube [2, 4]. For example, when a polymeric fluid is sheared, polymers are stretched and experience elastic and orientational stress. In the Doi-Edwards model, the stress on the polymer relaxes as it and the surrounding polymers return to their equilibrium lengths and conformations. At short times, or for small displacements, elastic relaxation is predicted to dominate, with the relevant time scale being the Rouse relaxation time $\tau_R$. For $t > \tau_R$ the polymers are predicted to relax by reptation, on the time scales up to the disengagement time $\tau_D$. Relaxation times have been measured in many rheological experiments but agreement with theory is not always good. Many efforts have been made to improve the theory by introducing various corrections such as contour length fluctuations, residual stretch relaxation, convective constraint release, etc. [2, 4, 12-14].

Here, we introduce a new experimental approach in which the forces imposed by entangled polymers to confine a single stretched DNA molecule are measured directly using optical tweezers [10, 15]}. We determine the form of the potential energy landscape restricting molecular displacement. We also characterize the time dependence of the forces by studying the dependence on displacement rate and the force relaxation following displacement.

The probe molecule was a 25.3 kbp DNA (8.4 μm) prepared by PCR as described [16]. It was embedded in a 1 mg/ml (~40 times the overlap concentration c* [17, 18]) solution of linear 115 kbp (38 μm) DNA, prepared by replication in *E. coli* as described [17]. Previous experiments confirm that we are in the well-entangled regime [8, 10].

The probe molecule was attached by each end to optically trapped microspheres (1.1 μm and 1.4 μm radii) and stretched with an applied tension of 10 pN (corresponding to a fractional extension ≅ 0.95) [19] as described [16, 20] (Fig. 1A&B). The surrounding molecules forming the constraints were allowed to relax. The optical traps were calibrated as described [20]. To map the confining field, the sample chamber was displaced relative to the trapped probe molecule using a piezoelectric nanopositioning stage (Mad City Labs). Displacements were made either transverse or parallel to the probe chain contour at constant velocity. The force along the axis of displacement was recorded at 1 kHz and low-pass filtered with a negative-exponential filter (0.1 sampling proportion in a 1 μm range) (Sigmaplot 6.0, SPSS, Inc.). Measurements were also done without the probe molecule to determine the contribution of the microspheres (~70%), and this force was subtracted from the total force in order to isolate the force acting on the probe molecule. Measurements were repeated 20 times at different locations in the sample chamber to verify reproducibility and accurately determine the average induced force. A small increase in $F_x$ was measured during *y*-displacements, indicating an increase in tension due to slight bowing of the probe molecule. However, for displacements up to 2 μm (those relevant in probing the tube potential) this effect was negligible ($\Delta F_x$ <5% of $\Delta F_y$), and could thus be ignored in our analysis of small displacement data.

Semidilute solutions of DNA molecules and numerous synthetic polymers have been shown to have universal dynamic and rheological properties when scaled according to blob theory [6]. Thus, we chose to scale our measured confining force by the correlation blob size of our entangled polymer solution. Using the expression $\zeta = R_G(c/c^*)^{-\nu/3\nu-1}$ [1], where $R_G$ is the radius of gyration [21] we calculate a correlation blob size of $\zeta \cong 52$ nm. By dividing the measured force by $L/\zeta$, where $L = 8.4$ μm is the probe length, we can calculate the confining force per blob or unit length.

According to traditional Doi-Edwards theory the tube radius is given by $a = (4/5(M_e/M)6R_G^2)^{1/2}$ where $M_e$ is the molecular weight between entanglements and $M$ is the molecular weight [2]. $R_G$ is determined from previous work [8, 21, 22] and $M_e$ is determined from bulk rheological measurements of the plateau modulus, $G_N^{(0)}$ via the relationship $M_e =$ $(4/5)cRT/G_N^{(0)}$, where $R$ is the gas constant and $T$ the temperature [2]. Based on recent measurements of $G_N^{(0)}$ with entangled DNA [8], and the prediction [2] that it is independent of $M$ and proportional to $c^2$, we estimate $G_N^{(0)} \cong 0.5$ Pa and calculate $a \cong 0.5$ μm.

Force measurements were made with transverse displacements ranging from zero up to 15.5 μm. For large displacements (>~3 μm), significant bowing of the probe molecule and the possibility of generating responses in the nonlinear or "strong flow" regime make the data more difficult to interpret. Because small length scales, on the order of the predicted tube radius $a \cong 0.5$ μm, are most relevant in probing the tube potential, we focus most of our attention on analyzing the small displacement (<2 μm) data.

According to reptation theory thermally diffusing chain segments reach the tube radius and thus "feel" the effects of the tube on the time scale of the equilibration time $\tau_e = a^4/24D_G R_G^2$ [2, 23]. Using previously determined values for $R_G$ and $D_G$ (diffusion coefficient) we calculate $\tau_e \cong 0.02$ s [8, 21, 22]. Therefore, the characteristic rate at which chain segments are predicted to move a distance $a$ via thermal motion is $a/\tau_e = 25$ μm/s. Thus, we made displacements at 25 μm/s, in addition to a range of higher and lower rates (0.01 to 65 μm/s).

For small displacements the force increased approximately linearly with distance (Hookean behavior)(Fig. 1D&E). The effective spring constant increased with increasing velocity. In contrast, negligible forces were induced in response to parallel displacements (Fig. 1C). A potential concern with parallel displacements is that the microsphere may partly disturb the surrounding chains through which the probe moves. However, the maximum displacement of 2.6 μm was much shorter than the 8.4 μm length of the probe molecule so ~70% of the probe was moving through surrounding chains that were undisturbed. Thus, we have shown by direct measurement that the collective intermolecular forces imposed by entangled polymers give rise to a tube-shaped confining field. By integrating the perpendicular force versus displacement we can determine the transverse confining potential per unit length $U(y)$ (Fig. 1F&G).

Zhou and Larson [9] calculated the tube potential $U(y)$ in simulations by determining the relative probability $P(y)$ that a monomer of an entangled polymer fluctuates a certain transverse distance $y$ from the primitive path and by using the relationship $P(y)=e^{-U(y)/kT}$. Although direct quantitative comparison with our results is not warranted because they simulated a melt rather than a solution, our findings are in qualitative agreement, as expected from blob theory. They find a time-dependent harmonic confining potential for short times, in accord with our findings. They also find that the tube radius calculated by conventional theory is similar to the characteristic length at which $U(y) \cong 1\ kT$ ($P(y) \cong 1/e$) at time $\tau_e$. Likewise, we can determine a characteristic tube radius at the displacement where $U(y) \cong kT$ (Fig. 2). At the relevant $a/\tau_e \cong 25$ μm/s rate we find $U = kT$ at ~0.8 μm, similar to the theoretically predicted tube radius of 0.5 μm.

While the conventional tube model assumes that the tube radius is time-independent, our measurements and the simulations of Zhou and Larson find a time-dependent potential [9]. Zhou and Larson have suggested that this effect can help to explain certain longstanding discrepancies between rheological data and theory. They calculate that the tube radius increases by ~10% on a time ranging from $\tau_e$ to $100\tau_e$. Similarly, our measured tube radius increases by ~20% with time increasing from $\tau_e$ to $50\tau_e$.

The longest predicted relaxation time in Doi-Edwards theory is the disengagement time $\tau_D = (18R_G^2/a^2)\tau_R \cong 40$ s, which is the characteristic time it takes a chain to completely diffuse out of its initial tube by reptation. On long times approaching $\tau_D$ we expect the tube-like constraints to weaken and eventually disappear. Indeed, at the lowest displacement rate in our experiment (0.1 μm/s), corresponding to ~$10a/\tau_D$, we find a much lower potential of ~0.2 $kT$ at the 0.8 μm tube radius. It rises to ~1 $kT$ for a displacement of ~2 μm, indicating ineffective tube confinement on this long time scale.

Our results for large displacements (>~3 μm) are more difficult to interpret due to the issues discussed above. We nevertheless present these findings (Fig. 1E&G), which may be of interest since the tube model has been applied in many cases to nonlinear states generated by strong flows [2, 4]. A notable feature in all three cases is that a decrease in $dF/dy$ was observed for $\Delta y > 7$ μm. This distance is close to the predicted primitive path length of the entangled DNA in the Doi-Edwards theory ($L_0 \cong (M/M_e)a \cong 9$ μm) [2], so this decrease might indicate displacement-induced slippage of entanglements off of the probe molecule. Such behavior may be related to the effect of "convective constraint release" believed to be important in understanding the non-linear response of polymeric liquids in steady shear flow [8].

Further information came from the measured force relaxation following both small (2.6 μm) and large (15.5 μm) displacements (Fig. 3). Inverse Laplace transform analysis [24] revealed three distinct decay times for all data sets except the 0.1 μm/s case, where only the two faster relaxation times were observed. The time constants and relative amplitudes were insensitive to the rate and size of the displacement. To accurately determine the time constants the averaged relaxation curves were fitted to a discrete sum of three decaying exponentials, obtaining $\tau_1 = 0.45 \pm 0.34$ s, $\tau_2 = 5.4 \pm 2.8$ s, and $\tau_3 = 34 \pm 5.8$ s. The relative amplitudes of these decay times were all roughly equal (each ~1/3 of the total).

These measurements probe the relaxation of the distorted molecules surrounding the probe molecule. According to Doi-Edwards theory [2], the shortest relaxation time is the Rouse time $\tau_R = 6R_G^2/3\pi^2 D_G$, on which a deformed polymer elastically relaxes back to its equilibrium primitive path length ($L_0$). We calculate $\tau_R \cong 0.6$ s for our system [8, 21, 22], in good agreement with the shortest relaxation time measured ($\tau_1$). Theory also predicts that elastic relaxation accounts for only 1/5 of the total relaxation, attributing the other 4/5 to disengagement from the tube [25]. However, we find a ~1/3 contribution for $\tau_1$. Higher frequency Rouse modes occurring at progressively shorter times are also predicted to contribute to the relaxation, however they are often neglected due to the short times over which they act [25]. It is possible that the larger amplitude we find is due to the contribution of unresolved higher order Rouse modes to $\tau_1$.

The longest predicted relaxation time, the disengagement time $\tau_D \cong 40$ s, is consistent with our measured value of $\tau_3$. The 1/3 contribution measured is much lower than the predicted 4/5, however we find a distinct intermediate decay time, $\tau_2 \cong 12\tau_1$, not predicted by Doi-Edwards theory, that also contributes to the relaxation. The observed amplitudes thus suggest that reptation is occurring on this intermediate time scale $\tau_2$ as well. A relaxation time similar to our measured $\tau_2$ was recently observed in nonlinear step shear experiments with polystyrene solutions [13, 26] and in flow deformation studies of DNA at 35c* [8]. A possible explanation for $\tau_2$ has been proposed by Mhetar and Archer [14]. During displacement, the tube diameter is proposed to shrink as its length is increased, prohibiting complete relaxation of chain extension to $L_0$ in time $\tau_R$. An additional relaxation mode with decay time $\tau_R < t < \tau_D$ is predicted whereby the residual stretch relaxes as the tube expands back to its equilibrium diameter. Reptation is also predicted to occur on this timescale, however complete disengagement from the tube only occurs after a time $\tau_D$.

Finally, we examine the dependence of the confining force on length and concentration of the surrounding chains. We made additional measurements using shorter 11 kbp DNA at 1 mg/ml (~7.5c*, $\zeta \cong 50$ nm) and 115 kbp DNA diluted to 0.1 mg/ml (~4c*, $\zeta \cong 300$ nm). The measured confining forces per unit length were similar for both cases and significantly lower than that for the 115 kbp, 1 mg/ml solution. With the most extreme deformation (15.5 μm at 65 μm/s) the maximum force was only ~50 fN for both cases compared with ~200 fN for the 115 kbp, 1 mg/ml solution. Also, the relaxations were notably different in that both were well described by only a single time constant, ~0.5 s for the shorter construct and ~0.6 s for the dilute solution, which are both close to the predicted value of $\tau_R$ as well as our measured fastest relaxation time $\tau_1$. These findings suggest that at these lower concentrations or lengths the molecules are not yet entangled so there is no effective confining tube.

**References**

[1] P. G. de Gennes, *Scaling concepts in polymer physics* (Cornell University Press, Ithaca, N.Y., 1979).
[2] M. Doi and S. F. Edwards, *The theory of polymer dynamics* (Oxford University Press, Oxford, 1986).
[3] R. G. Larson, *The Structure and Rheology of Complex Fluids* (Oxford University Press, Oxford, 1998).
[4] T. C. B. McLeish, Advances in Phys **51**, 1379 (2002).
[5] R. H. Colby and M. Rubinstein, Macromolecules **23**, 2753 (1990).
[6] Y. Heo and R. G. Larson, J. of Rheology **49**, 1117 (2005).
[7] K. Kremer and G. S. Grest, J Chem Phys **92**, 5057 (1990).
[8] R. E. Teixeira, A. K. Dambal, D. H. Richter, et al., Macromolecules **40**, 2461 (2007).
[9] Q. Zhou and R. G. Larson, Macromolecules **39**, 6737 (2006).
[10] T. T. Perkins, D. E. Smith, and S. Chu, Science **264**, 819 (1994).
[11] D. E. Smith, T. T. Perkins, and S. Chu, Phys Rev Lett **75**, 4146 (1995).
[12] J. P. Montfort, G. Marin, and P. Monge, Macromolecules **17**, 1551 (1984).
[13] L. A. Archer, J. Sanchez-Reyes, and Juliani, Macromolecules **35**, 10216 (2002).
[14] V. Mhetar and L. A. Archer, J Non-Newt Fluid Mech **81**, 71 (1999).
[15] S. B. Smith, Y. Cui, and C. Bustamante, Science **271**, 795 (1996).
[16] D. N. Fuller, G. J. Gemmen, J. P. Rickgauer, et al., Nucleic Acids Res **34** (2006).
[17] S. Laib, R. M. Robertson, and D. E. Smith, Macromolecules **39**, 4115 (2006).
[18] R. M. Robertson and D. E. Smith, Macromolecules **40**, 3373 (2007).
[19] C. Bustamante, J. F. Marko, E. D. Siggia, et al., Science **265**, 1599 (1994).
[20] J. P. Rickgauer, D. N. Fuller, and D. E. Smith, Biophys J **91**, 4253 (2006).
[21] R. M. Robertson, S. Laib, and D. E. Smith, Proc Nat Acad Sci USA **103**, 7310 (2006).
[22] J. S. Hur, E. S. G. Shaqfeh, H. P. Babcock, et al., J. Rheology **45**, 421 (2001).
[23] M. Putz, K. Kremer, and G. S. Grest, Europhys Lett **49**, 735 (2000).

[24] W. H. Press, S. A. Teukolsy, W. T. Vetterling, et al., *Numerical Recipes in C* (Cambridge University Press, Cambridge, 1992).
[25] A. E. Likhtman and T. C. B. McLeish, Macromolecules **35**, 6332 (2002).
[26] J. Sanchez-Reyes and L. A. Archer, Macromolecules **35**, 5194 (2002).

**Figure Captions**

**Fig 1.** **(A)** Schematic diagram of the experiment. A single DNA molecule is held stretched between two optically trapped microspheres and suspended in a concentrated solution of entangled DNA. **(B)** Reptation models postulate that collective intermolecular interactions give rise to a tube-shaped confining field (dashed lines). We measure the confining force per unit length ($F_x$ and $F_y$) in response to an imposed displacement $\Delta x$ or $\Delta y$ (see text). **(C)** Average force induced by a transverse displacement $\Delta y$ at 13 µm/s (gray points) compared with that induced by a parallel displacement $\Delta x$ at 65 µm/s (black points). Arrows mark the maximum displacements. The inset graph shows the displacement profiles. **(D)** $F_y$ vs. $\Delta y$ at velocities of 65 µm/s (red), 25 µm/s (blue), 13 µm/s (green), 0.52 µm/s (cyan), and 0.10 µm/s (orange). **(E)** Results with a large displacement ($\Delta y = 15.5$ µm). **(F)** Confining potential energy per unit length $U(y)$ determined by integration of force data in plot (D). The dashed line indicates the theoretically predicted tube radius. **(G)** $U(y)$ determined by integration of force data in plot (E).

**Fig 2.** Relative probability for a transverse displacement *y* from the primitive path determined by $P(y) = e^{-U(y)/kT}$ where $U(y)$ is the confining potential (Fig 1F). Colors indicate different displacement velocities as defined in Fig 1. Tube radius defined as the distance where $P(y) = 1/e$, indicated by the dashed line. Measured tube radii are: 0.65 µm, 0.82 µm, 0.94 µm, 1.0 µm and 1.8 µm for the 65 µm/s, 25 µm/s ($a/\tau_e$), 13 µm/s, 0.52 µm/s, and 0.10 µm/s respectively.

**Fig 3.** Force relaxation following transverse displacements of 2.6 or 15.5 µm. The colors indicate different displacement velocities as defined in Fig 1. The solid lines are fits to sums of decaying exponentials (see text). Decay constants in seconds for the 2.6 µm ramp: 0.82, 11.8 for 0.10 µm/s; 0.1, 4.3, 38 for 0.52 µm/s; 0.6, 4.1, 40 for 13 µm/s; and 0.51, 3.9, 39 for 25 µm/s. Decay times for the 15.5 µm ramp: 0.05, 4.6, 28 for 0.52 µm/s; 0.20, 4.5, 28 for 13 µm/s; and 0.86, 4.8, 29 for 65 µm/s.

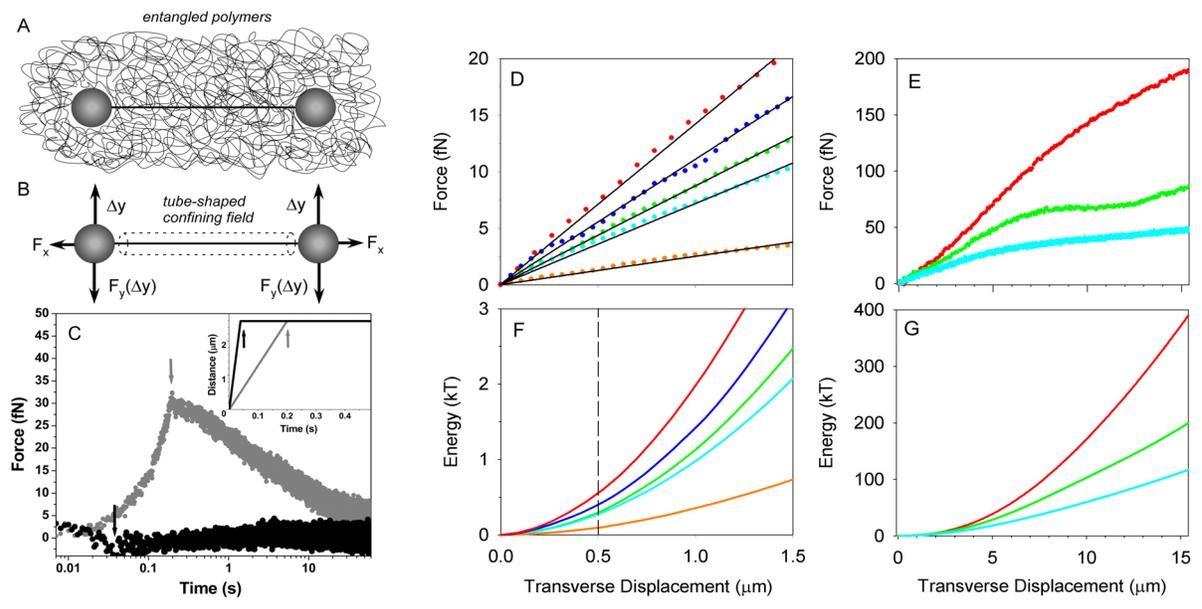

**Fig. 1**

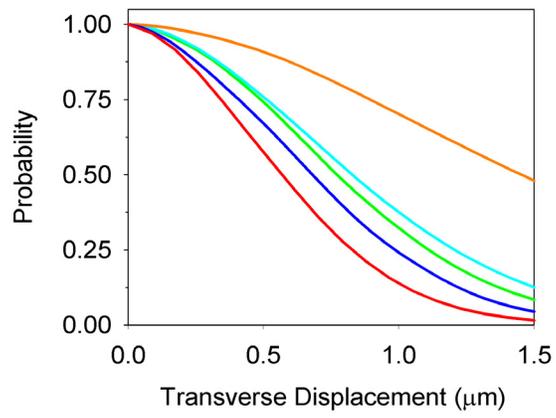

**Fig. 2**

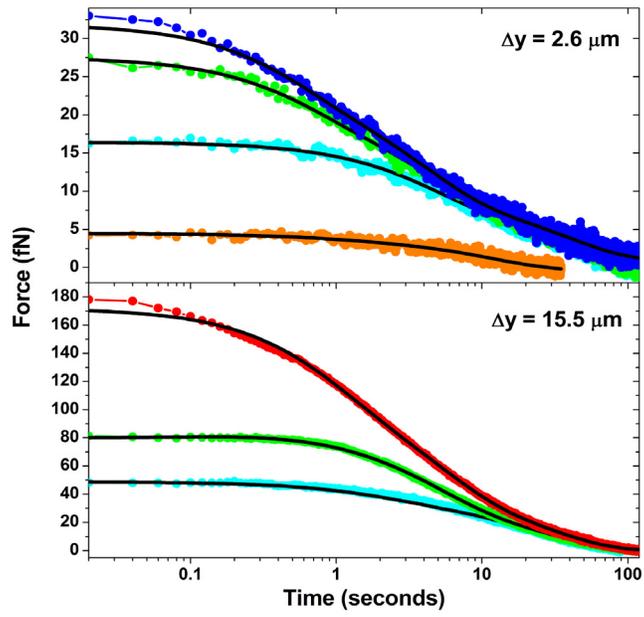

**Fig. 3**